\documentclass[%
preprint,
 amsmath,amssymb,
 aps,
]{revtex4-2}

\usepackage{dcolumn}
\usepackage{amssymb,amsmath}
\usepackage{hyperref}
\usepackage{graphicx}
\usepackage{bm}
\usepackage{orcidlink}
\usepackage[mathlines]{lineno}
\usepackage{cancel}

\begin{document}

\title{Emergent Gravity Completion in Quantum Field Theory, and Affine Condensation in Open and Closed Strings}

\author{Durmu{\c s} Demir\orcidlink{0000-0002-6289-9635}}
\email{durmus.demir@sabanciuniv.edu}
\homepage{http://myweb.sabanciuniv.edu/durmusdemir/}
\affiliation{Sabanc{\i} University, Faculty of Engineering and Natural Sciences, 34956 {\.I}stanbul, T{\"u}rkiye}

\date{\today}

\begin{abstract}
The ultraviolet cutoff on a quantum field theory can be interpreted as a condensate of the affine curvature such that while the maximum of the affine action gives the power-law corrections, its minimum leads to the emergence of gravity. This mechanism applies also to fundamental strings as their spinless unstable ground levels can be represented by the scalar affine curvature such that open strings (D-branes) decay to closed strings and closed strings to finite minima with emergent gravity. Affine curvature is less sensitive to massive string levels than the tachyon, and the field-theoretic and stringy emergent gravities take the same form. It may be that affine condensation provides an additional link between the string theory and the known physics at low energies.
\end{abstract}

\maketitle

\tableofcontents

\section{Introduction}
\label{sec:intro}
The SM is a quantum field theory (QFT) of the electromagnetic, weak, and strong interactions \cite{wein0}. It leaves out gravity. It inherently belongs to the flat spacetime due to difficulties with the quantization of the curved metric \cite{quant-gravity} and difficulties also with the carriage of QFTs to curved spacetime \cite{qfields-curved,weinx2,weinx3}. Nevertheless, unlike the full QFTs like the SM, flat spacetime effective QFTs can have a particular affinity with curved spacetime since they are nearly-classical field theories obtained by integrating out high-energy quantum fluctuations \cite{wein0,wein1} in the sense of both the Wilsonian and one-particle-irreducible effective actions \cite{polchinski, burgess, renormalization}.

Viewed as an effective QFT, the SM must have a physical UV validity limit (not a regulator), say $\Lambda_\wp$ \cite{wein1}. It is the scale at which a UV completion sets in. It can lie at any scale above a TeV according to the null results from the LHC searches \cite{lhc}. It breaks Poincare (translation) symmetry explicitly so that the loop momenta $\ell^\mu$ lie in the range $-\Lambda_\wp^2 \leq \eta_{\mu\nu} \ell^\mu \ell^\nu \leq \Lambda_\wp^2$
in flat spacetime with Minkowski metric $\eta_{\mu\nu}$. Then, loop corrections to scalar mass-squareds contain terms proportional to $\Lambda_\wp^2$. Likewise, the  vacuum energy involves  $\Lambda_\wp^4$ and $\Lambda_\wp^2$ terms \cite{wein1,veltman}. In addition to these unnatural UV-sensitivities \cite{wein1}, all of the gauge bosons acquire mass-squareds proportional to $\Lambda_\wp^2$, and these loop-induced masses break gauge symmetries explicitly \cite{gauge-break1,gauge-break2}. What would be more natural than to use the Higgs mechanism \cite{Higgs1, Higgs2, Higgs3}  to restore the gauge symmetries? Impeding this proposal is the radical difference between the intermediate vector boson mass (Poincare-conserving) \cite{wein2}  and the loop-induced gauge boson mass (Poincare-breaking) \cite{dad2023}. Indeed, in the former, in accordance with the Poincare conservation,  vector boson mass is promoted to a scalar field, which leads to the usual Higgs mechanism \cite{Higgs1, Higgs2, Higgs3}. In the latter, due to the Poincare-breaking nature of $\Lambda_\wp$, it is necessary to find a Poincare-breaking Higgs field and employ the Higgs mechanism accordingly. In this regard, we construct a gauge symmetry restoration mechanism in which affine curvature \cite{affine1,affine2} comes into play as the Higgs field promoting the UV cutoff $\Lambda_\wp$, and condenses to the usual metrical curvature in the minimum of the metric-affine action \cite{affine3,affine4} such that unnatural UV sensitivities get defused, gauge symmetries are restored, the emergence of gravity is enabled, and new particles are brought about that do not have to couple to the SM directly \cite{dad2023,dad2021,dad2019,dad2016}. This affine condensation mechanism 
gives a bottom-up UV completion of the QFT with emergent gravity. We shall construct it in detail in Sec. \ref{sec:bottom-up}. 

In string theory, the ground levels of both the open (D-brane) and closed strings are spinless imaginary-mass states. They are unstable. They are readily represented by a scalar tachyon as part of the corresponding string field theory \cite{string-tachyon1,string-tachyon2}. There is, however, a problem in that the mass of the tachyon, in absolute magnitude, is of a similar size as those of the massive string levels, and the tachyon can therefore hardly give a low-energy effective field theory description of the massless string spectrum. In regard to this problem, we realize that actually the affine curvature in Sec. \ref{sec:bottom-up} can give a proper description of the string ground level because it contains the usual massless graviton as part of the low-energy effective theory. As we show in Sec. \ref{sec:top-down}, affine condensation leads to a picture in which open strings (D-brane) decay into closed strings and closed strings decay into not the vacuity but a minimum in which gravity emerges just as in Sec. \ref{sec:bottom-up}. This top-down approach gives a stringy completion of the QFTs. 

In Sec. \ref{sec:conclude}, we give a comparative discussion of the bottom-up and top-down approaches. In particular, we conclude that affine condensation may be the way string theory is linked to the known physics at low energies. In this section, we discuss also few future prospects. 

\section{Bottom-Up: Emergent Gravity Completion of Quantum Field Theory}
\label{sec:bottom-up}
In this section, following a bottom-up approach, we shall discuss the completion of the QFT in the UV via emergent gravity. In essence, gravity will emerge in a way that restores the gauge symmetries via the condensation of the affine curvature such that the affine curvature itself enters the game as a Higgs-like field promoting the loop-induced gauge boson masses namely the UV cutoff.

\subsection{Effective QFT}
Let us consider a generic renormalizable field theory of quantum fields $\psi$ (the SM fields plus new particles) for generality. In flat spacetime, the effective QFT capturing  the physics of the full QFT at low energies $\mu \ll \Lambda_\wp$ is described by the effective action \cite{dad2023}
\begin{eqnarray}
S[\eta, \psi; \log\mu, \Lambda_\wp^2] = S_{tree}[\eta, \psi] +  \delta S_{log}[\eta,  \psi; \log\mu] + \delta S_{pow}[\eta, \psi; \log\mu, \Lambda_\wp^2]
\label{eff-ac}
\end{eqnarray}
in extension  of the dimensional regularization \cite{dimreg1,dimreg2} to QFTs with a physical UV cutoff $\Lambda_\wp$ \cite{detached}. In the effective action, the tree-level action  $S_{tree}$  sets symmetries, field spectrum, and interactions in the QFT. The logarithmic correction  $\delta S_{log}$ involves the renormalization scale $\mu$ (not the UV cutoff $\Lambda_\wp$) and follows the structure of $S_{tree}$. The power-law correction 
\begin{eqnarray}
\delta S_{pow} = \int d^4x \sqrt{-\eta}\, \Big\{-c_O \Lambda_\wp^4 - 2 M_O^2 \Lambda_\wp^2 - c_\phi \Lambda_\wp^2 \phi^\dagger \phi + c_V \Lambda_\wp^2 {\rm tr}\left[V^\mu \eta_{\mu\nu}  V^\nu\right] \Big\} 
\label{Sp}
\end{eqnarray}
involves the scalar fields $\phi$ and the gauge fields $V_h\mu$ with the color trace ${\rm tr}[\dots]$. Here, the loop-induced factors
\begin{eqnarray}
c_O&=& \frac{(n_b-n_f)}{64 \pi^2}
\label{cO}
\end{eqnarray}
and 
\begin{eqnarray}
M_O^2 &=& \frac{{\rm str}\!\left[M^2\right]}{64\pi^2} 
\label{MO}
\end{eqnarray}
set the power-law corrections to the vacuum energy, with $n_b$ ($n_f$) being the total number of bosonic (fermionic) degrees of freedom in the QFT, and $M^2$ denoting the mass-squared matrix of all the QFT fields, with the supertrace ${\rm str}[M^2]=\sum_J (-1)^{2 J} (2 J +1) M_J^2$ over the particle spin $J$. The loop factor $c_\phi$ controls the quadratic  corrections to scalar mass-squareds ($c_\phi({\rm Higgs})\approx 3 h_t^2/4\pi$). The loop factor $c_V$, on the other hand, rules the breaking explicitly of the gauge symmetry symmetries ($c_V({\rm gluon})=21 g_3^2/16\pi^2$ and $c_V({\rm hypercharge})=39 g_3^2/32\pi^2$ ) \cite{dad2021,dad2019}. 

\subsection{Taking effective QFT to curved spacetime}
The matter fields $\psi$ in (\ref{eff-ac}) are feebly-fluctuating quantum fields (nearly classical), and this ensures that the effective action $S[\eta]$ can be smoothly carried into spacetime of a curved metric $g_{\mu\nu}$ via general covariance \cite{covariance} 
\begin{eqnarray}
   \eta_{\mu\nu} \rightarrow g_{\mu\nu}\,,\;\; \partial_\mu \rightarrow \nabla_\mu  
   \label{covariance}
\end{eqnarray}
in which $\nabla_\mu$ is the covariant derivative  with respect to the Levi-Civita connection 
\begin{eqnarray}
    {}^g\Gamma^\lambda_{\mu\nu} = \frac{1}{2} g^{\lambda\rho} \left( \partial_\mu g_{\nu\rho} + \partial_\nu g_{\rho\mu} - \partial_\rho g_{\mu\nu}\right)
    \label{LC}
\end{eqnarray}
so that $\nabla_\alpha g_{\mu\nu}=0$. But, for the metric $g_{\mu\nu}$ to be curved, the effective QFT in curved spacetime must involve the metrical curvature, like, for example, the Ricci curvature $R_{\mu\nu}({}^g\Gamma)$. In classical field theories,  spacetime metric can be made dynamical (curved) by simply adding requisite curvature terms (such as the Einstein-Hilbert term) with appropriate bare constants. In effective field theories, this is not possible. The reason is that effective QFTs have all their couplings generated or corrected by loop contributions, and adding any curvature term by hand contradicts the renormalized QFT structure. Indeed, the introduction of bare terms means that the curvature sector of the original QFT was left unrenormalized while the matter sector was renormalized. In the face of this contradiction, the curvature must arise from within the flat spacetime effective action (\ref{eff-ac}) not with some bare constants but with the loop factors $c_i$. In other words, curvature must arise via the deformations of the existing loop corrections in $\delta S_{pow}$. Possible deformations are set by the commutator $[\nabla_\mu,\nabla_\nu]$ as it generates curvature via its actions $V^\mu[\nabla_\nu,\nabla_\mu] V^\nu = V^\mu R_{\mu\nu}({}^g\Gamma) V^\nu$ on the gauge fields $V^\mu$ and ${\overline{\psi}}\gamma^\mu \gamma^\nu [\nabla_\mu,\nabla_\nu] \psi = - (1/2) g^{\mu\nu}R_{\mu\nu}({}^g\Gamma) {\overline{\psi}} \psi$ on the fermions $\psi$. This dimension-5 fermion contribution is irrelevant. Thus, deformations can occur only in the gauge sector. In fact, the structure of $V^\mu R_{\mu\nu}({}^g\Gamma) V^\nu$ suggests that the object to be deformed is the gauge boson mass term $c_V \Lambda_\wp^2 {\rm tr}\left[V^\mu \eta_{\mu\nu}  V^\nu\right]$. To see this, it proves useful to start with the flat spacetime  null deformation 
\begin{eqnarray}
{\widetilde S}[\eta] &=& S[\eta]+\int d^4x \sqrt{-\eta} c_V {\rm tr}\left[V^\mu\!\left( \left[D_\mu,D_\nu\right] -iV_{\mu\nu}\right)\! V^\nu\right]
\nonumber\\ &=& S[\eta]
\label{iden-flat}
\end{eqnarray}
where the added deformation term consists of the field strength tensor $V_{\mu\nu}$ of the gauge fields defined as  $[D_\mu,D_\nu] V^\nu = iV_{\mu\nu} V^\nu$ such that $D_\mu = \partial_\mu + i V_\mu$ is the gauge-covariant derivative, with $V^\mu$ being a vector in the adjoint.
Obviously, the deformation integral in (\ref{iden-flat}) vanishes identically, and therefore the flat spacetime effective action $S[\eta]$ remains untouched, ${\widetilde S}[\eta] \equiv S[\eta]$. But, under the general covariance map in (\ref{covariance}), this null deformation  gives rise to a  curved spacetime  action 
\begin{eqnarray}
{\widetilde S}[g] &=& S[g] + \int d^4x \sqrt{-g} c_V {\rm tr}\left[V^\mu \!\left( \left[{\mathcal D}_\mu,{\mathcal D}_\nu\right] -iV_{\mu\nu}\right)\! V^\nu\right]\nonumber\\
&=&  S[g] - \!\!\int\! d^4x \sqrt{-g} c_V {\rm tr}\left[V^\mu R_{\mu\nu}({}^g\Gamma) V^\nu\right]
\label{iden-curved}
\end{eqnarray}
whose second line develops the sought-for curvature tensor as generated by the commutator
$[{\mathcal D}_\mu,{\mathcal D}_\nu] V^\nu = \left(-R_{\mu\nu}({}^g\Gamma) + iV_{\mu\nu}\right) V^\nu$ of the curved spacetime covariant derivative  ${\mathcal D}_\mu = \nabla_\mu + i V_\mu$. The curvature in (\ref{iden-curved}) ensures that spacetime is curved. But there is no gravity! There is no gravity because the Einstein-Hilbert term is missing in (\ref{iden-curved}). This imperative term, proportional to the curvature scalar $g^{\mu\nu} R_{\mu\nu}({}^g\Gamma)$, has to emerge from within the curved spacetime effective action (\ref{iden-curved}). 

Our goal is to find out how the Einstein-Hilbert term emerges from within the curved spacetime effective action ${\widetilde S}[g]$. For this, it proves useful to contrast the two mass terms: $M_{I}^2 {\rm tr}\!\left[I_\mu I^\mu\right]$ and $c_V \Lambda_\wp^2 {\rm tr}\!\left[V_\mu V^\mu\right]$. In the former, $M_{I}$ is the Poincare-conserving mass of an intermediate vector boson $I_\mu$ (like the W/Z bosons) and thus it can be promoted as 
\begin{eqnarray}
M_I^2 {\rm tr}\!\left[I_\mu I^\mu\right] \longmapsto (I_\mu S)^\dagger I^\mu S \subset (D_\mu S)^\dagger D^\mu S
\label{promote-higgs}
\end{eqnarray}
to a Poincare-conserving Higgs scalar $S$ for restoring the gauge symmetry with gauge-covariant derivative $D_\mu=\partial_\mu +i I_\mu$ \cite{Higgs1,Higgs2,Higgs3,wein2}. In the latter, however, $\Lambda_\wp^2$ breaks the Poincare symmetry, and so it can be promoted to only a Poincare-breaking field (not the scalar $S$) for restoring the gauge symmetry \cite{dad2023,dad2021,dad2019,dad2016}. In general, Poincare-breaking fields are expected to be related to the spacetime curvature since, in a general second-quantized field theory with no presumed symmetries, Poincare symmetry is known to emerge if the Poincare-breaking terms are identified with curvature \cite{fn}. 
Nonetheless, the Poincare-breaking field promoting $\Lambda_\wp^2$ must be insensitive to the metric as it must remain nonzero in flat and curved spacetimes. One field that satisfies both of these conditions is the Ricci curvature ${\mathbb{R}}_{\mu\nu}(\Gamma)$ of a general affine connection $\Gamma^\lambda_{\mu\nu}$. Indeed, $\Gamma^\lambda_{\mu\nu}$  is independent of the metric $g_{\mu\nu}$ but tends to ${}^g\Gamma^\lambda_{\mu\nu}$ in (\ref{LC}) by affine dynamics, and, as a result, ${\mathbb{R}}_{\mu\nu}(\Gamma)$ tends to  $R_{\mu\nu}({}^g\Gamma)$. This dynamical nearing ensures that ${\mathbb{R}}_{\mu\nu}(\Gamma)$ is the sought-for Poincare-breaking field, and the UV cutoff $\Lambda_\wp$ can therefore be promoted as 
\begin{eqnarray}
\Lambda_\wp^2 g_{\mu\nu}
\longmapsto  {\mathbb{R}}_{\mu\nu}(\Gamma) 
\label{MV-map-new}
\end{eqnarray}
for restoring the gauge symmetries \cite{dad2023,dad2021,dad2019,dad2016}. This map parallels the promotion in (\ref{promote-higgs})
of the intermediate vector boson masses $M_I$ to scalars $S$  for restoring the gauge symmetries \cite{wein2}. It makes the curved spacetime effective action ${\widetilde S}[g]$ in (\ref{iden-curved}) a metric-Palatini gravity theory \cite{affine4,affine5}
\begin{eqnarray}
{\widetilde S}[g,\Gamma] &=& S_{tree}[g, \psi] +  \delta S_{log}[g,  \psi; \log\mu]\nonumber\\ &+& \int d^4x \sqrt{-g}\, \left\{-\frac{c_O}{16} {\mathbb{R}}^2 
-\frac{M_O^2}{2}{\mathbb{R}}   -\frac{c_\phi}{4} \phi^\dagger\phi\, {\mathbb{R}} + c_V {\rm tr}\left[V^\mu\!\left( {\mathbb{R}}_{\mu\nu} - R_{\mu\nu}\right) V^\nu\right]
\right\}
\label{S-metric-affine}
\end{eqnarray}
in which ${\mathbb{R}} = g^{\mu\nu}{\mathbb{R}}_{\mu\nu}(\Gamma)$ is the affine scalar curvature. Its coefficient $M_O^2$ must be positive and much bigger than any QFT scale since metric-Palatini dynamics will eventually turn it to the fundamental scale of gravity \cite{affine2,affine3,affine4}
\begin{eqnarray}
M_O
= M_{Pl}
\label{MPl}
\end{eqnarray}
and ascribe it this way to a new physical role compared to its definition in (\ref{MO}) and role in (\ref{Sp}) in the flat spacetime. 

\subsection{Emergent gravity from affine condensation}
\label{Gauge symmetry-restoring}
Owing to its effective nature (nearly classical), stationary points of the metric-Palatini action (\ref{S-metric-affine}) are expected to give the extremal points corresponding to physical field configurations. Then, keeping the matter sector in its vacuum state ($\langle V_\mu\rangle =0$, $\langle f\rangle =0$, $\langle \phi \rangle/ M_{Pl}\approx 0$), extremal values of $\Gamma$ are derived from the stationarity condition 
\begin{eqnarray}
    \frac{\delta {\widetilde S}[g,\Gamma]}{\delta \Gamma^\lambda_{\mu\nu}} = 0 \Longrightarrow {}^\Gamma\nabla_\lambda\! \left({\mathbb{Q}}^{1/3}g^{\mu\nu}\right) = 0
    \label{eom}
\end{eqnarray}
in which the covariant derivative 
${}^\Gamma\nabla_\lambda$ is that of  $\Gamma$, and 
\begin{eqnarray}
    {\mathbb{Q}}= \frac{M_{Pl}^2}{2}+ 
    \frac{c_O}{8} {\mathbb{R}}
    \label{Q-tens}
\end{eqnarray}
equals the variation of $-{\widetilde S}[g,\Gamma]$ with ${\mathbb{R}}$ in the vacuum.

One solution of the motion equation  (\ref{eom}) is just ${\mathbb{Q}}=0$. And it gives a constant affine curvature scalar $-{4 M_{Pl}^2}/{c_O}$ from  (\ref{Q-tens}). This is the extremal curvature that leaves the action (\ref{S-metric-affine}) stationary. It must be equal to $4 \Lambda_\wp^2$ since the affine curvature ${\mathbb{R}}$ is after all the ``Higgs field" promoting $\Lambda_\wp^2$ as in (\ref{MV-map-new}), and its extremal value must therefore regenerate the UV cutoff. But for this regeneration, $-4 M_{Pl}^2/c_O$ must be positive, and this can happen only if
\begin{eqnarray}
   c_O<0 \Longrightarrow n_b < n_f 
   \label{cO-kucuk}
\end{eqnarray}
revealing that nature has more fermions than bosons. This negative $c_O$ makes the ${\mathbb{R}}$--potential in (\ref{S-metric-affine}) maximum at the extremum  $-4 M_{Pl}^2/c_O>0$. Thus, ${\mathbb{Q}}=0$ solution with $c_O<0$ leads to the extremal  affine curvature scalar
\begin{eqnarray}
{\mathbb{R}}_{\rm max} = - \frac{4 M_{Pl}^2}{c_O} = 4 \Lambda_\wp^2 
\label{Rmax}
\end{eqnarray}
at which $-{\widetilde S}[g,\Gamma]$ attains its maximum. At this maximal curvature, it reduces to the curved spacetime action in (\ref{iden-curved}) and brings back, therefore, all the power-law UV corrections in (\ref{Sp}). This maximum is full of problems.

One other solution of the motion equation  (\ref{eom}) occurs with ${\mathbb{Q}}\neq 0$ and leads to the minimal configuration
\begin{eqnarray}
    (\Gamma_{\rm min})^\lambda_{\mu\nu}= {}^g\Gamma^\lambda_{\mu\nu} + \frac{1}{6{\mathbb{Q}}}\! \left(\nabla_\mu  \delta^\lambda_\nu +  \nabla_\nu  \delta^\lambda_\mu -\nabla^\lambda g_{\mu\nu} \right)\!{\mathbb{Q}} 
    \label{gamma-soln}
\end{eqnarray}
as the maximal configuration was already identified in (\ref{Rmax}). 
Expanding (\ref{gamma-soln}) up to ${\mathcal{O}}\!\left(M_{Pl}^{-4}\right)$ terms, 
the extremal affine curvature is found to be  
\begin{eqnarray}
  ({\mathbb{R}}_{\rm min})_{\mu\nu}= R_{\mu\nu}({}^g\Gamma) -\frac{c_O}{24 M_{Pl}^2} \!\left(\Box  g_{\mu\nu} + 2 \nabla_\mu \nabla_\nu \right)\!R
  \label{R-soln-exp}
\end{eqnarray}
where $R=g^{\mu\nu}\!R_{\mu\nu}({}^g\Gamma)$ is the metrical curvature scalar. In this minimum, the $c_V$ part of (\ref{S-metric-affine}) reduces to
\begin{eqnarray}
\int d^4x \sqrt{-g}\, c_V {\rm tr}\left[V^\mu\!\left({\rm zero} +  {\mathcal{O}}\!\left(M_{Pl}^{-2}\right)\right) V^\nu\right]
\label{fark-gauge} 
\end{eqnarray}
which is seen to vanish up to ${\mathcal{O}}\!\left(M_{Pl}^{-2}\right)$ terms. This nullification of the gauge boson mass terms ensures that all the gauge symmetries are restored in the minimum in (\ref{R-soln-exp}) up to doubly Planck-suppressed terms. 

\begin{figure}[ht!]
\centering 
\includegraphics[scale=0.94]{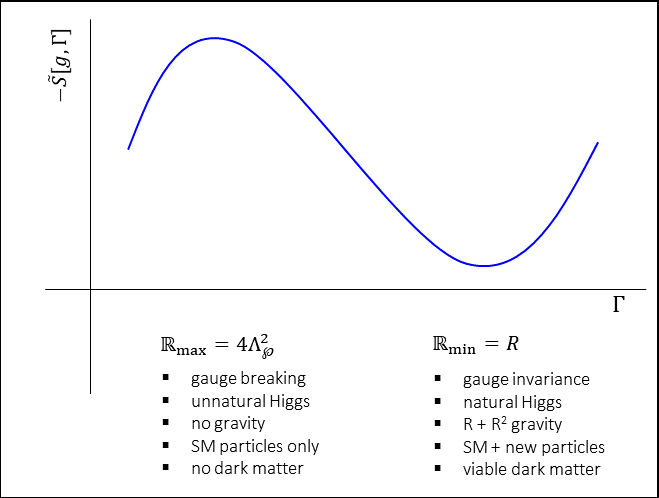} 
\caption{A schematic plot of $-{\widetilde S}[g,\Gamma]$ contrasting its maximum and minimum. Its minimum leads to symmergent gravity.}
\label{fig1}
\end{figure}

In the minimum \eqref{R-soln-exp}, the metric-Palatini effective action (\ref{S-metric-affine}) gives rise to, up to 
${\mathcal{O}}\!\left(M_{Pl}^{-2}\right)$ terms, a contemperature of dimensionally-regularized QFT and $R+R^2$ gravity theory
\begin{eqnarray}
S[g] = S_{tree}[g, \psi] +  \delta S_{log}[g,  \psi; \log\mu] + \int d^4x \sqrt{-g} \left\{-\frac{M_{Pl}^2}{2}R - \frac{c_O}{16} R^2 -\frac{c_\phi}{4} \phi^\dagger\phi R \right\}
\label{QFT+GR}
\end{eqnarray}
up to doubly Planck-suppressed terms, including the remainder in (\ref{fark-gauge}). This resultant action describes the emergence of the curvature sector via the promotion in (\ref{MV-map-new}) of the UV cutoff to affine curvature and condensation in (\ref{R-soln-exp}) of the affine curvature to the metrical curvature. It turns out that the affine curvature condensation gives rise to both the gauge and gravitational interactions. In essence, the action $S[g]$ is the curved spacetime image of the flat spacetime effective action (\ref{eff-ac}), and lays a gauge symmetry restoring emergent gravity framework, which we abbreviate as {\it symmergent gravity} \cite{dad2023,dad2021,dad2019}. Symmergence is the physics in the minimum. Its item-by-item comparison in Fig. \ref{fig1} with the physics at the maximum reveals that the QFT has transited from the problem-full maximum in (\ref{Rmax})  to the physically viable minimum in (\ref{R-soln-exp}) via the affine dynamics in (\ref{eom}).

Symmergence has important physics implications. Firstly, the gravitational scale $M_{Pl}$ comes out wrong in the SM given in its definition in (\ref{MO}) via (\ref{MPl}). So,  new particles with heavy bosons are necessary, but their couplings to the SM are not simply because they do not need any non-gravitational couplings to SM fields to saturate the mass sum rule in (\ref{MO}). LHC searches imply $\Lambda_\wp > 1\ {\rm TeV}$ \cite{lhc}, and this puts an upper bound of $n_f-n_b < 3.75\times  10^{33}$ from (\ref{Rmax}). In flat spacetime, there is no scale like the Planck mass, so $\Lambda_\wp$ can well be trans-Planckian, and in that case, one finds $n_f-n_b < 630$ for $\Lambda_\wp > M_{Pl}$.

Secondly, symmergence allows the Higgs and other possible light scalars to remain light. Indeed, symmergence completes the QFT by converting the power-law UV corrections in (\ref{Sp}) to curvature terms in (\ref{S-metric-affine}) so light scalars can be destabilized only by large logarithms in $\delta S_{log}$ of heavy new fields $\Psi$. But,  the logarithmic shift in the Higgs mass 
\begin{eqnarray}
\delta m_h^2 \propto \lambda_{h\Psi} M_{\Psi}^2 \log \frac{M_{\Psi}^2}{\mu^2}
\label{scal-mass}
\end{eqnarray} 
does not have to destabilize the SM since  Higgs coupling $\lambda_{h\Psi}$ to new fields $\Psi$ is allowed to be small in symmergence. In fact, $\lambda_{h\Psi}$ is allowed to even vanish with no harm to the workings of the symmergence. In general, couplings of the size $\lambda_{h \Psi}\!\lesssim \! m_h^2/m_\Psi^2$ stabilize $\delta m_h^2$ and render the Higgs boson natural \cite{natur-symm}. Also important is that the new fields can form a natural dark sector with none to weak couplings to the SM. They can even form a fully
decoupled black sector in agreement with the current dark matter searches \cite{dm1,dm2,natur-symm}. In this regard, dark/black stars and galaxies become a possibility.

Thirdly, symmergence implies that the Universe starts flat. Gravity emerges afterward from the quantum fluctuations of the matter fields. The vacuum energy from $\delta S_{log}$ in (\ref{QFT+GR})
\begin{eqnarray}
\delta V= \frac{1}{32\pi^2}  {\rm str} \left[M^4 \left(1-\frac{3}{2}\log \frac{M^2}{\mu^2}\right)\!\right]
\label{vac-en}
\end{eqnarray}
is of Planckian size as implied by the definition in (\ref{MPl}) of the Planck scale, and gives rise to a solution $R(g)\sim M_{Pl}^2$ via the Einstein field equations from (\ref{QFT+GR}). The maximum and minimum in Fig. \ref{fig1} then get split in energy by $\Lambda_\wp - M_{Pl}\sim \Lambda_\wp$ so that the gravity emerges within a time about $\Lambda_\wp^{-1}$. This symmergent evolution can be described by an FRW Universe with a quasi-linear scale factor such that the $\delta V$ in (\ref{vac-en}) sets the initial value of the scale factor \cite{eric1,eric2,eric3}. This FRW description holds initially and gets modified in time with higher-derivative contributions from the quadratic curvature term. This symmergent cosmology sharply contrasts quantum cosmology \cite{qc} in that the Universe starts flat (no gravity) in the former and curved (quantum gravity) in the latter. Also, the classical gravity phase is attained in a time around $\Lambda_\wp^{-1}$ in the former and $M_{Pl}^{-1}$ in the latter. 

Finally, the symmergent gravity action (\ref{QFT+GR}) gives an intertwined description of the renormalized QFT and classical gravity. The two sectors are compatible since quantum fluctuations have been integrated out in flat spacetime, and gravity has emerged upon it. In spite of this concord, the two sectors are still incompatible in the face of small remnant fluctuations of the quantum fields. Nevertheless, given the Einstein field equations following from  (\ref{QFT+GR}), fluctuations in quantum fields can be translated as stochastic fluctuations in the spacetime metric and, thanks to the loss of reversibility this way, the two sectors can be regarded to attain a certain degree of compatibility \cite{weinx2,weinx3}.

\section{Top-Down: Emergent Gravity Devocation of String Instability}
\label{sec:top-down}
Flatness at the start ensures that the progenitor of the whole enchilada can be structured and understood as a quantum object. One such object would be an open string of length scale $\Lambda_\wp^{-1}$ ending on D-branes \cite{string1,string2}. This is because open strings with Dirichlet boundary conditions break the Poincare symmetry. But non-supersymmetric open and closed strings do fatefully decay because their ground states are imaginary-mass, spin-zero states lying below even the massless states like the photon \cite{string-tachyon1}. Indeed, for a string with tension $T=\Lambda_\wp^2$, the ground level acquires a mass $M^2 = -2\pi \Lambda_\wp^2$ and $M^2 = -8\pi \Lambda_\wp^2$ for open and closed strings, respectively. Energetically, therefore, open strings and the D-branes they are attached to are expected to decay into closed strings \cite{string-tachyon7}.  

The decays and interactions of strings are studied in the framework of the equivalent string field theory \cite{string-tachyon2,string-tachyon11}. In this regard, the tachyonic scalars like the Higgs field \cite{Higgs1, Higgs2, Higgs3,wein2} are tailor-made for the spinless, imaginary-mass string ground level. In fact, the Schrodinger equation for the imaginary-mass string ground state is isomorphic to the Klein-Gordon equation for a classical tachyonic scalar \cite{string-tachyon2,string-tachyon3}. But there is a serious problem with this tachyonic string field theory. There is a problem because the mass of the string ground level, in absolute magnitude, is at the same scale as the second, third, and even higher excited levels of the string, and construction of a low-energy effective field theory for the tachyon deadlocks as it mixes with the massive excited states of the string \cite{string-tachyon2,string-tachyon1}. Indeed, as a toy model collecting results of different constructions \cite{string-tachyon2, string-tachyon4, string-tachyon5, Witten1992, Witten1985,string-tachyon2-1,string-tachyon2-2}, the minimal tachyon action
\begin{eqnarray}
S[\eta,{\mathbb{T}}]=\int d^D x \sqrt{-\eta} \left\{ \frac{1}{2} \eta^{\mu\nu} \partial_\mu {\mathbb{T}}\partial_\nu {\mathbb{T}} + \pi \Lambda_\wp^2 {\mathbb{T}}^2 +  {\tilde{c}}_O {\mathbb{T}}^4 + {\tilde{c}}_V {\rm{tr}}\left[V_\mu V^\mu\right] {\mathbb{T}}^2 + {\mathcal{L}}_{\rm mix}({\mathbb{T}},\chi_{\geq 2})\right\}
    \label{tachyon-condense}
\end{eqnarray}
inevitably contains an interaction term ${\mathcal{L}}_{\rm mix}({\mathbb{T}},\chi_{\geq 2})$ with the second and higher string levels $\chi_{\geq 2}$. The ${\tilde{c}}_V$ term couples ${\mathbb{T}}$ to the massless vectors $V_\mu$ in the first excited level of the open string. In the vacuum $\langle V_\mu \rangle = 0$ so that $-S[\eta,{\mathbb{T}}]$ is maximized at ${\mathbb{T}}_{\rm max}=0$ and minimized at ${\mathbb{T}}^2_{\rm min}=-\Lambda_\wp^2/2{\tilde{c}}_O$ provided that ${\tilde{c}}_O<0$ and provided also that  ${\mathcal{L}}_{\rm mix}({\mathbb{T}},\chi_{\geq 2})$  is negligible. The transition from ${\mathbb{T}}_{\rm max}$ to ${\mathbb{T}}_{\rm min}$ 
 gives a dynamical representation of the open string (D-brane) instability. The vacuum energy drops from zero at the maximum to about  $-\Lambda_\wp^D$ in the minimum, and the minimum contains only closed strings, with no open string excitations left \cite{string-tachyon6,string-tachyon2,string-tachyon3}. Also these closed strings decay into final states not containing closed string excitations, particularly the gravity \cite{string-tachyon8,string-tachyon9,string-tachyon10}. These extrema and decays are possible if  $|{\mathcal{L}}_{\rm mix}({\mathbb{T}},\chi_{\geq 2})|$ is negligibly small but this smallness is not possible. It is in this sense that the tachyon seems to fail in forming a low-energy effective field theory for the string.

\begin{figure}[ht!]
\centering 
\includegraphics[scale=0.78]{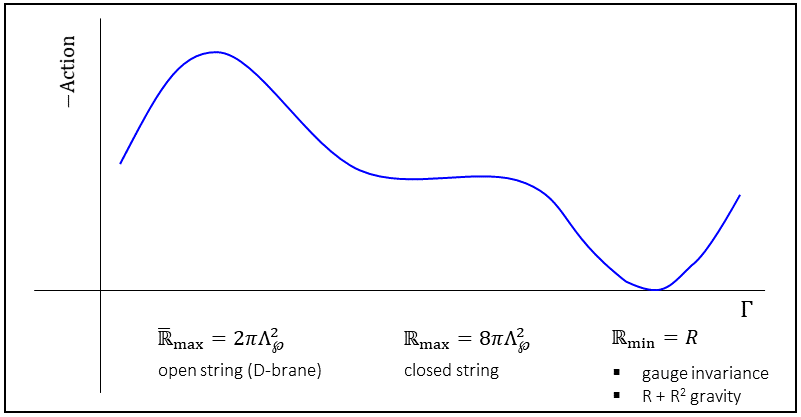} 
\caption{Open strings (D-branes) end on closed strings and closed strings end on a finite minimum with emergent gravity and gauge interactions. Structurally, the endpoint is the symmergent gravity.}
\label{fig2}
\end{figure}

The difficulty with the tachyon entails a natural question: Can the string ground level be represented by a massless unstable scalar? If yes, can the massless string excitations develop a low-energy effective action comprising gravity and matter? The answer is yes. Actually, the answer is symmergence. It is so because the affine scalar curvature in symmergence can well be the unstable scalar representing the string ground level. To see this, one first observes that a general affine connection $\Gamma^\lambda_{\mu\nu}$ transforms as a tensor (connection) in flat (curved) spacetime and has, in general, nothing to do with the metric tensor. Its curvature, the Ricci curvature ${\mathbb{R}}_{\mu\nu}(\Gamma)$ for instance, remains a genuine tensor field in both the flat and curved spacetimes \cite{dad2023}. In fact, mimicking the tachyon action in (\ref{tachyon-condense}) and remaining parallel to the symmergent metric-affine action in (\ref{S-metric-affine}), affine curvature can be attributed the action 
\begin{eqnarray}
S_{OS}\left[\eta,{\overline{\mathbb{R}}}\right] = \int d^Dx \sqrt{-\eta}\, \left\{-\frac{{\widehat{c}}_O}{16} {\overline{\mathbb{R}}}^2 
-\frac{{\widehat{M}}_O^2}{2}{\overline{\mathbb{R}}} + {\widehat{c}}_{OV}  {\rm tr} \left[V_\mu V^\mu\right] {\overline{\mathbb{R}}} \right\}
\label{affine-condense-flat}
\end{eqnarray}
in the $D$-dimensional flat spacetime of the open string (D-brane). Here, ${\overline{\mathbb{R}}} = \eta^{\mu\nu} {\mathbb{R}}_{\mu\nu} (\Gamma)$ is a mass dimension-two scalar acting as the Higgs-like field promoting the mass-squared of the string ground level. 
In essence, the affine action in (\ref{affine-condense-flat}) is a replacement for the tachyon action in (\ref{tachyon-condense}) since, after all, it involves a the scalar ${\overline{\mathbb{R}}}$ in place of the tachyon ${\mathbb{T}}$. They differ from each other by the absence of the mixing term  ${\mathcal{L}}_{\rm mix}({\mathbb{T}},\chi_{\geq 2})$, and the reason for this is that ${\overline{\mathbb{R}}}$ is not a canonical massive scalar and, as will be seen in the sequel, ${\mathbb{R}}_{\mu\nu}(\Gamma)$ consists of massless modes. The massless vector $V_\mu$ in the first excited level of the open string must be present in the low-energy effective action and, for this to happen,  $V_\mu$ must couple to ${\overline{\mathbb{R}}}$ and such a coupling is given at the lowest order either by the ${\widehat{c}}_{OV}$ term in (\ref{affine-condense-flat}) or by $V^{\mu} {\mathbb{R}}_{\mu\nu}(\Gamma)V^\nu$ term in symmergence since gauge kinetic term is independent of the connection. In the vacuum, $\langle V_\mu \rangle = 0$ and $-S_{OS}[\eta,{\overline{\mathbb{R}}}]$ gets extremized at ${\overline{\mathbb{R}}}_{\rm max}=-4 {\widehat{M}}_O^2/{\widehat{c}}_O$. This extremum becomes a maximum for ${\widehat{M}}_O^2>0$ and ${\widehat{c}}_O<0$ and being an unstable ${\overline{\mathbb{R}}}$ configuration it gives a representation of the open string ground level provided that ${\widehat{c}}_O$ and ${\widehat{M}}_O^2$ satisfy the relation ${\overline{\mathbb{R}}}_{\rm max}=-4 {\widehat{M}}_O^2/{\widehat{c}}_O=2\pi \Lambda_\wp^2$. As was already discussed while deriving (\ref{Rmax}) in Sec. \ref{Gauge symmetry-restoring},  ${\overline{\mathbb{R}}}={\overline{\mathbb{R}}}_{\rm max}$ is the solution of the equation ${\mathbb{Q}}= {\widehat{M}}_O^2/2+ 
    {\widehat{c}}_O {\overline{\mathbb{R}}}/8=0$ where ${\mathbb{Q}}$ is what controls the motion equation
  ${}^\Gamma\nabla_\lambda\! \left({\mathbb{Q}}^{1/3}\eta^{\mu\nu}\right) = 0$. The other solution with ${\mathbb{Q}}\neq 0$ would lead to ${\overline{\mathbb{R}}}=0$ in flat spacetime as revealed by equation (\ref{R-soln-exp}) in Sec. \ref{Gauge symmetry-restoring},  and this means complete destruction of the open string (D-brane). But before this destruction occurs closed strings come into play as decay products. The massless dilatonic scalar $\phi$ and the massless spin-2 field $h_{\mu\nu}$ in the first excited level of closed strings modify the massless spectrum. The main effect of $h_{\mu\nu}$ is to generate a curved metric $\eta_{\mu\nu} \rightarrow \eta_{\mu\nu}+h_{\mu\nu} =g_{\mu\nu}$ in the same spirit as the general covariance map in (\ref{covariance}). Inclusion of these massless fields modifies the open string action \eqref{affine-condense-flat} to give it the closed string form  
\begin{eqnarray}
S_{CS}[g,{\mathbb{R}}] =  \int d^Dx \sqrt{-g}\, \left\{-\frac{{\widehat{c}}_C}{16} {\mathbb{R}}^2 
-\frac{{\widehat{M}}_C^2}{2}{\mathbb{R}} - \frac{{\widehat{c}}_{\phi}}{4} \phi^\dagger \phi {\mathbb{R}} + \left({\widehat{c}}_{CV} {\mathbb{R}} + {\Breve{c}}_{CV} R\right) {\rm tr} \left[V_\mu V^\mu\right] \right\}
\label{affine-condense-curved}
\end{eqnarray}
in which ${\mathbb{R}}=g^{\mu\nu} {\mathbb{R}}_{\mu\nu}(\Gamma)$ is the affine scalar curvature, $R=g^{\mu\nu}R_{\mu\nu} ({}^g\Gamma)$ is the metrical scalar curvature, and ${\widehat{c}}_C$, ${\widehat{M}}_C^2$, ${\widehat{c}}_{CV}$ are analogs of the open string constants ${\widehat{c}}_O$, ${\widehat{M}}_O^2$, ${\widehat{c}}_{OV}$ in (\ref{affine-condense-flat}). The new ${\Breve{c}}_{CV}$ term brings in the metrical curvature $R$ and ensures this way that  $g_{\mu\nu}$ is a curved and invertible metric. The new ${\widehat{c}}_{\phi}$ term, on the other hand, couples ${\mathbb{R}}$ to the massless scalars $\phi$. There is no ${\mathbb{R}} R$ term because of the ghosts  \cite{ghost1,ghost2,ghost3}. The instability of the closed string is the instability of the entire spacetime hence its endpoint would be vacuous of gauge, gravity, and matter in the absence of ${\mathbb{R}}$, and in this sense, metrical terms like $R$, $R^2$ and $R\phi^\dagger\phi$ are avoided as they can induce gravity by themselves.  In the vacuum in which $\langle V_\mu \rangle = 0$ and $\langle \phi^\dagger \phi\rangle \ll {\widehat{M}}_C^2$, the action $-S_{CS}[g,{\mathbb{R}}]$ gets extremized at ${\mathbb{R}}_{\rm max}=-4 {\widehat{M}}_C^2/{\widehat{c}}_C= 8\pi \Lambda_{\wp}^2$ assuming, for simplicity, similar tensions for the closed and open strings.  This extremum becomes a maximum for ${\widehat{M}}_C^2>0$ and ${\widehat{c}}_C<0$ and, as an unstable configuration, gives a representation of the closed string ground level. In the language of \eqref{Rmax}, this is the ${\mathbb{Q}}=0$ solution with ${\mathbb{Q}}= {\widehat{M}}_{C}^2/2+ 
    {\widehat{c}}_{C} {\mathbb{R}}/8$. The other solution with ${\mathbb{Q}}\neq 0$, as revealed by equation (\ref{R-soln-exp}) in Sec. \ref{Gauge symmetry-restoring}, leads to a nontrivial solution 
\begin{eqnarray}
{\mathbb{R}}_{\rm min} = R + {\mathcal{O}}\big({\widehat{M}}_C^{-2}\big)
\label{curv-min}
\end{eqnarray}
 in curved spacetime of the metric $g_{\mu\nu}$. This solution sets up a minimum at $\Gamma=\Gamma_{\rm min}$ following the maximum at $\Gamma=\Gamma_{\rm max}$. Its replacement  in the gauge part of (\ref{affine-condense-curved}) leads to the restoration of gauge symmetries up to ${\widehat{M}}_C^{-2}$ order if ${\widehat{c}}_{CV}$ 
and  ${\Breve{c}}_{CV}$ are related as 
\begin{eqnarray}
{\widehat{c}}_{CV}=-{\Breve{c}}_{CV}    
\end{eqnarray}
and in this minimum of the restored gauge symmetry, the rest of the action (\ref{affine-condense-curved}) reduces to 
\begin{eqnarray}
S_{CS}[g] = S_{QFT}[g,\psi] + \int d^Dx \sqrt{-g} \left\{- \frac{{\widehat{M}}_C^2}{2} R - \frac{{\widehat{c}}_C}{16} R^2 -\frac{{\widehat{c}}_\phi}{4} \phi^\dagger\phi R \right\}
\label{QFT+GR+st}
\end{eqnarray}
after explicating the effective action $S_{QFT}[g,\psi]$ of the massless scalar, vector and spinor string excitations $\psi$. The structure of gravity and matter in four dimensions depends on how these massless excitations propagate in the $(D-4)$--dimensional extra space. In case they depend only on the four non-compact dimensions,  the $S_{CS}[g]$ conforms to the symmergent gravity action $S[g]$ in (\ref{QFT+GR}) with the following parametric relations
\begin{eqnarray}
 M_{Pl}^2 = V_{D-4} \times {\widehat{M}}_C^2\;,\, c_O= V_{D-4} \times {\widehat{c}}_C\;,\, c_\phi=  {\widehat{c}}_\phi
 \label{rels}
\end{eqnarray}
where $V_{D-4}$ is the volume of the extra-dimensional space. This affine curvature condensation mechanism is summarized in Fig. \ref{fig2}. As the figure shows, open strings (D-brane) end with closed strings, and closed strings end with not the vacuity but the stringy symmergent gravity in (\ref{QFT+GR+st}). In this regard, symmergence seems to be the key mechanism in both bottom-up and top-down directions.  It links string theory to the known physics at low energies via the relations (\ref{rels}). It possesses the salient features below:
\begin{itemize}
    \item Firstly, fermions can be coupled to the open string action (\ref{affine-condense-flat}) or the closed string action (\ref{affine-condense-curved}) via the affine connection $\Gamma^\lambda_{\mu\nu}$ in their spin connections. This fermionic extension is not meant to introduce supersymmetry though imaginary-mass ground levels exist also in realistic superstrings \cite{string-tachyon2,string-tachyon2-0}.  
    
    \item Secondly, the stringy symmergent action (\ref{QFT+GR+st}) is an $R+R^2$ gravity theory like the symmergent action (\ref{QFT+GR}). It contains a scalar graviton having a mass around $\Lambda_\wp$. This mode, like the tachyon, can mix with the massive string modes $\chi_{\geq 2}$. It cannot therefore be part of the low-energy spectrum. The action (\ref{QFT+GR+st}) contains a massless spin-2 graviton, too.  This massless graviton and the other massless fields in $S_{QFT}[g,\psi]$ form a consistent low-energy effective field theory. It is in this sense that the affine condensation does better than the tachyon condensation.
    
    \item Thirdly, the string ground-level actions in (\ref{affine-condense-flat}) and (\ref{affine-condense-curved}) are essentially toy models illustrating the main physics implications. They are nevertheless robust in that they are structured by eyeing the tachyon action (\ref{tachyon-condense}). They give therefore information about what kind of string field-theoretic structures are to be considered in describing the ground-level instability \cite{string-tachyon2, string-tachyon4, string-tachyon5, Witten1992, Witten1985,string-tachyon2-1,string-tachyon2-2}. 
    
    \item Lastly, in this novel picture of affine curvature condensation, string theory is linked to physics at low energies (the SM plus additional particles) via not only the compactification of the extra dimensions but also the emergence of gravity and gauge symmetries. It seems that emergence could be the way string theory accesses the real world.
\end{itemize}

\section{Conclusion}
\label{sec:conclude}
This work is composed of two parts. In the first part, we followed a bottom-up approach to complete effective QFTs in the UV via affine curvature condensation. This has led to the restoration of gauge interactions, the emergence of gravity, and the renormalization of the underlying QFT. This mechanism, the symmergent gravity, intertwines gravity and quantum fields, with the introduction of new particles that do not have to couple to the SM directly, and with the allowance to suppress couplings to heavy fields for enabling stabilization of the scalar masses. In this picture, gravity emerges upon flat spacetime effective QFT, and, in a cosmological setting, the Universe starts out flat where the gauge forces and gravity take shape in a time set by the UV cutoff. This symmergent gravity framework  can be probed via cosmic inflation \cite{irfan} and reheating; black holes \cite{bh1,bh2,bh3,bh4,bh5,bh6,bh7}, wormholes and neutron stars; and collider \cite{lhc-deney} and beyond-the-collider \cite{natur-symm}  experiments.  

In the second part, we followed a top-down approach to reveal that the instabilities in the ground levels of the open and closed strings are better modeled by affine scalar curvature rather than the tachyon.  We have provided an affine curvature condensation picture of the open string (D-brane) decay followed by the closed string decay. Here, the main novelty is that the closed string ground state is described by the symmergent gravity, with model parameters coming from strings, not the QFT.  It may be that the string theory is linked to the known physics at low energies not only by the compactification of extra dimensions but also by the emergence of gravity along with the gauge interactions. This top-down approach gives a stringy completion of the QFTs. 

For future prospects, one research direction would be the explicit construction of the string field theory that leads to the stringy metric-affine dynamics as a generalization presumably of the earlier works \cite{string-tachyon2, Witten1992, Witten1985,string-tachyon2-1,string-tachyon2-2}. Under symmergence, such a string field-theoretic construction has the potential to determine the string parameters from the low-energy experimental bounds (via for example the quadratic-curvature coefficient $c_O$ involving $n_B-n_F$). Another direction would be the analysis of the very early Universe where the usual quantum gravity phase is replaced by flat spacetime \cite{eric1,eric2,eric3}. Yet another direction would be a test of the symmergence (including the stringy construction) in astrophysical compact objects like neutron stars. There are actually various directions to investigate. In summary, with the realization especially of the stringy completion, symmergent gravity can form a new framework for investigating various astrophysical, cosmological, and collider phenomena \cite{dad2023,dad2021,dad2019,dad2016}.

\section*{Acknowledgements}
I am grateful to Eric Ling for bringing Ref. \cite{eric1} to my attention. I thank Orfeu Bertolami, Patrick Das Gupta, Canan Karahan, Eric Ling,  Sebastian Murk, Ali {\"O}vg{\"u}n, Beyhan Puli{\c c}e, Ozan Sarg{\i}n, and Kai Schwenzer for fruitful discussions on different aspects of this work. 
\bibliography{References.bib}

\nolinenumbers

\end{document}